\documentclass[aps,twocolumn,showpacs,superscriptaddress]{revtex4}

\usepackage{graphicx}
\usepackage{amsmath}

\def\Dth{\Delta\theta}
\def\Dp{\Delta p}

\begin{document}

\author{Pierre de Buyl}
\affiliation{Chemical Physics Theory Group, Department of Chemistry, University of Toronto, Toronto, Ontario, M5S 3H6 Canada}
\affiliation{Center for Nonlinear Phenomena and Complex Systems, Universit{\'e} Libre de Bruxelles
(U.L.B.), Code Postal 231, Campus Plaine, B-1050 Brussels, Belgium}
\author{David Mukamel}
\affiliation{Department of Physics of Complex Systems, Weizmann Institute of Science, Rehovot 76100, Israel}
\author{Stefano Ruffo}
\affiliation{Dipartimento di Energetica and CSDC, Universit{\`a} di Firenze and CNISM, via s. Marta 3, 50139 Firenze, Italy}
\affiliation{Laboratoire de Physique de l'\'{E}cole Normale Sup\'{e}rieure de Lyon, Universit\'{e} de Lyon, CNRS, 46 All\'{e}e d'Italie, 
69364 Lyon c\'edex 07, France}

\title{Self-consistent inhomogeneous steady states in Hamiltonian mean field dynamics}

\begin{abstract}
Long-lived {\it quasistationary states}, associated with stationary stable solutions of the
Vlasov equation, are found in systems with
long-range interactions. Studies of the relaxation time in a model
of $N$ globally coupled particles moving on a ring, the Hamiltonian
Mean Field model (HMF), have shown that it diverges as $N^\gamma$
for large $N$, with $\gamma \simeq 1.7$ for some initial conditions with homogeneously
distributed particles. We propose a method for identifying {\it exact inhomogeneous steady
states} in the thermodynamic limit, based on analysing models of
uncoupled particles moving in an external field. For the HMF model, we show numerically
that the relaxation time of these states diverges with $N$ with the
exponent $\gamma \simeq 1$. The method, applicable to other models with
globally coupled particles, also allows an exact evaluation of the
stability limit of homogeneous steady states. In some cases it provides a
good approximation for the correspondence between the initial condition and the
final steady state.
\end{abstract}

\maketitle

\section{Introduction}
\label{sec:introduction}

Systems with long-range interactions exhibit unusual thermodynamic
and dynamical properties
\cite{reviews}.
These systems are characterized by two-body interaction potentials
which decay at large distances $r$ as $r^{-\alpha}$, with $\alpha$
smaller than the spatial dimension $d$. With such potentials these
systems are not additive and, as a result, they display
characteristic features such as ensemble inequivalence, negative
specific heat, temperature jumps and broken ergodicity.

Studies of the dynamics of a class of models have shown that some
initial macroscopic states display slow relaxation to equilibrium.
Starting from these initial states the system evolves on a short
timescale towards a state that is not the one predicted by
equilibrium statistical mechanics. The lifetime of this {\it
quasistationary state} (QSS) diverges algebraically with system
size. For a finite system one eventually relaxes towards thermal
equilibrium \cite{yamaguchi_et_al_physica_a_2004}. The
full nonlinear dynamical relaxation process remains to be fully
characterized.

A prototypical model for which this dynamical behavior has been
explored in detail is the Hamiltonian Mean Field model
(HMF)\cite{HMF}, which describes the motion on a ring
of $N$ fully coupled particles. Its Hamiltonian is
\begin{equation}
\label{eq:HMF}
\mathcal{H} = \sum_{i=1}^N \frac{p_i^2}{2} + \frac{1}{2N} \sum_{i,j=1}^N
\left( 1 -\cos (\theta_j-\theta_i)\right)
\end{equation}
where $-\pi< \theta_i\leq \pi$ is the position of the $i$-th
particle and $p_i$ is its conjugated momentum. The model has a
second order phase transition at the energy $u_c={\cal H}/N=3/4$ at
which the order parameter $\mathbf{m}=(m_x,m_y)=\sum_i(\cos
\theta_i,\sin \theta_i)/N$ changes from zero (above $u_c$) to non
zero (below $u_c$). This order parameter is referred to as the
magnetization vector.

For this mean-field model, as for other long-range interacting
systems, the time evolution of the single-particle distribution
function $f(\theta,p,t)$ is conveniently studied in the large $N$
limit by the Vlasov equation \cite{nicholson}
\begin{equation}
\label{vlasov}
\frac{\partial f}{\partial t} + p \frac{\partial f}{\partial \theta} -
\frac{\partial V[f](\theta,t)}{\partial \theta}
\frac{\partial f}{\partial p} = 0~,
\end{equation}
where
\begin{equation}
\label{potential} V[f](\theta,t)=\iint d \theta' dp'
f(\theta',p',t)\left(1-\cos (\theta'-\theta)\right)
\end{equation}
is the effective potential seen by the particles.

It has been proposed that linearly stable stationary solutions of
the Vlasov equation are associated with QSS of the large but finite
$N$ system~\cite{yamaguchi_et_al_physica_a_2004}. This has been discussed in detail 
for {\it homogeneous},
$\theta$ independent, steady states where $\mathbf{m}=0$ .
Studying {\it inhomogeneous} ($\mathbf{m} \neq 0$) steady states in this
class of models is less straightforward \cite{inhomogeneous}. Exploring these
stationary states, their stability and the relaxation process
for quasistationarity is a challenging goal. 

The stationary solutions $f_S(\theta,p)$ of the Vlasov equation are
functions of the integrals of motion and thus, in one dimension,
they are generically of the form
\begin{equation}
  \label{eq:stationary}
  f_S(\theta,p)=F(h(\theta,p)) \textrm{ with } h(\theta,p)=\frac{p^2}{2}+V[f_S](\theta).
\end{equation}
Here $h(\theta,p)$ is the single-particle energy. The function $F$
has to satisfy a self-consistency condition obtained by inserting
(\ref{eq:stationary}) into (\ref{potential}). However, this is satisfied by a
wide class of functions $F$, reflecting the fact that
the Vlasov equation has many stationary states. In this regard, a widely
used concept is that of BGK modes \cite{BGK,nicholson}. An application
of this method to the HMF model has been proposed in \cite{Yamaguchi_2011}. We will here 
propose a different approach.

In this paper we introduce a simple method for constructing inhomogeneous
steady states by analyzing the dynamics of models of non-interacting particles moving in an
external potential. Relaxation in such systems has already been
considered in Refs.~\cite{pomeau_2007,leoncini_et_al_epl_2009,de_buyl_mukamel_ruffo_inprep},
however it has not been used to derive analytically inhomogeneous steady states
of the interacting particles systems. 
A different approach yielding approximate inhomogeneous steady
states in systems with long-range interactions has previously been introduced
by Lynden-Bell \cite{lynden-bell_1967} and subsequently applied in a variety of models 
\cite{barre_et_al_pre_2004,antoniazzi_et_al_prl_2007,
levin,de_buyl_mukamel_ruffo_inprep}. Our method yields exact steady states and enables
one to evaluate their stability limit. It also provides a procedure to relate in an approximate 
way an initial distribution to the steady state to which it evolves. 
For the sake of clarity, the method is applied to the HMF model,
although it could be readily formulated in a more general context.

The paper is organized as follows. In Section~\ref{sec:uncoupled} we introduce the uncoupled
particles model and we derive the invariant measure. Section~\ref{sec:HMF} presents the application
of our method to the HMF model. In Section~\ref{sec:stability} we derive the exact stability
limit of the homogeneous waterbag state. Section~\ref{sec:conclusions} is devoted to some
conclusions.

\section{The uncoupled particles model}
\label{sec:uncoupled}

\begin{figure}[ht]
  \centering
  \includegraphics[width=\linewidth]{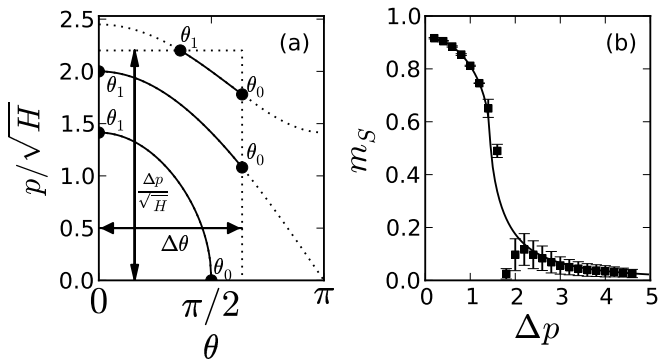}
  \caption{a) Graphical representation of the computation of Eq.~(\ref{P(e<H)}). The full lines
  represent the domain on which the integral is performed in order to compute $P_\epsilon$ for values
  of the energy $\epsilon=0$, $H$ and $2H$. The boundaries $\theta_0$ and $\theta_1$ are displayed
  for each energy. b) Magnetization in the steady state $m_s$ vs. $\Dp$ for an initial waterbag state with $\Delta \theta=1$.
  The full line is the theoretical prediction and the points are the results of numerical integrations of the
  Vlasov equation for the HMF model in Eqs.~(\ref{vlasov}) and (\ref{potential}). Error bars quantify fluctuations 
  of $m_s$ in the steady state.}
  \label{fig:pen-integ-wb}
\end{figure}

In order to find the stationary states of the HMF model, we consider the dynamics of an ensemble
of {\it uncoupled particles} moving in a fixed external field $H$
pointing, say, in the $x$-direction. The energy of a single particle
is given by
\begin{equation}
  \label{eq:energy}
  \epsilon(\theta,p)=\frac{p^2}{2} -H \cos \theta~.
\end{equation}
For an arbitrary function $F(\epsilon(\theta,p))$ to be a steady
state of the interacting model (\ref{eq:HMF}), $H$ has to satisfy the
self-consistency condition
\begin{equation}
  \label{eq:cons1}
  H=m_x=\iint d \theta dp F(\epsilon(\theta,p)) \cos \theta \quad ; \quad m_y=0~.
\end{equation}
In the following we suppress the subscript $x$. Any $F$ that
satisfies these conditions yields an exact stationary solution of the HMF
model. In order to relate an initial distribution to the steady
state to which it evolves, we consider an initial distribution of
particles $f_0(\theta,p)$. The dynamics of the uncoupled particles
model (\ref{eq:energy}) is such that particles in a given energy shell
$[\epsilon,\epsilon+d\epsilon]$ keep moving inside that shell,
eventually reaching a homogeneous distribution within it. 
As a result, the system attains the following steady state distribution
\begin{equation}
P(\theta,p)=\frac{\iint d\theta' dp' f_0(\theta',p') \, \delta \left(\epsilon(\theta',p')-\epsilon(\theta,p)\right)}
{\iint d\theta' dp' \, \delta \left( \epsilon(\theta',p')-\epsilon(\theta,p) \right)}~.
\end{equation}
For simplicity, we present below the analysis for the often studied {\it waterbag} initial
condition
\begin{equation}
  \label{eq:wb}
  f_0(\theta,p) =
  \left\{\begin{array}{l l l}
     \left(4 \Dth \Dp \right)^{-1} &,\textrm{ for } |\theta|\leq \Dth \textrm{ and } |p|\leq \Dp~,\cr
      0  &, \textrm{ otherwise.}
    \end{array}\right.
\end{equation}
In order to evaluate $P(\theta,p)$, it is convenient to first
consider the energy distribution $P_{\epsilon}(\epsilon)$. For the
waterbag initial state (\ref{eq:wb}) it is given by
\begin{equation}
P_{\epsilon}(\epsilon)=\frac{1}{4 \Dth \Dp} \int d\theta \int_{-\Dp}^{\Dp} dp
~\delta(\frac{p^2}{2}-H\cos\theta -\epsilon) ~.
\end{equation}
Carrying out the integral over $p$, one gets
\begin{equation}
P_{\epsilon}(\epsilon)=\frac{1}{2 \Dth \Dp} \int d \theta
~\frac{1}{\sqrt{2(\epsilon+H\cos\theta)}}~,
\end{equation}
for $-H \leq \epsilon \leq \Dp^2/2-H\cos \Dth$ and zero outside this range.
The integration over $\theta$ needs to be done in the domain
enclosed by the waterbag initial condition, namely
\begin{equation}
0 \le \epsilon+H\cos\theta \le \frac{\Dp^2}{2}~,
\end{equation}
Thus,
\begin{equation}
\label{P(e<H)} 
P_{\epsilon}(\epsilon)=\frac{\sqrt{2}}{ 2 \Dth\Dp}
\int_{\theta_1}^{\theta_0} d \theta \frac{1}{\sqrt{(\epsilon+H\cos\theta)}}
\end{equation}
where $\theta_0$ and $\theta_1$ satisfy (see Fig.~\ref{fig:pen-integ-wb}a)

\begin{equation}
  \label{eq:pen-theta0}
  \theta_0 = \left\{\begin{array}{l l}
      \arccos(-\epsilon/H) &, \textrm{ for } -H < \epsilon < -H \cos \Dth \\
      \Dth &, \textrm{ for } \epsilon \geq -H \cos \Dth~,
    \end{array}\right.
\end{equation}
\begin{equation}
 \label{eq:pen-theta1}
 \theta_1 = \left\{\begin{array}{l l}
     0 &,\textrm{ for } -H < \epsilon \leq \Dp^2/2 - H \\
     \arccos(\frac{\Dp^2/2-\epsilon}{H}) &, \textrm{ for } \Dp^2/2 - H \leq \epsilon <  \\
     &\qquad\ \Dp^2/2 - H \cos \Dth \\
     \Dth &,\textrm{ for } \epsilon \geq \Dp^2/2 - H \cos \Dth~.
 \end{array}\right.
\end{equation}

In the steady state, the $(\theta, p)$ distribution is such that, for
any given energy, all the microstates corresponding to that energy are
equally probable. The boundaries on $(\theta,p)$ imposed by the initial waterbag
are no longer valid. Thus, the steady state distribution $P(\theta, p)$
may be expressed as
\begin{equation}
\label{eq:distr1}
P(\theta, p)=\frac{1}{4\Dth\Dp}\frac{P_\epsilon(\epsilon(\theta,p))}{Q_{\epsilon}(\epsilon(\theta,p))}
\equiv\bar{P}_\epsilon(\epsilon(\theta,p))~,
\end{equation}
where $Q_{\epsilon}(\epsilon(\theta,p))$ is given by $P_\epsilon(\epsilon(\theta,p))$, evaluated by Eq.~(\ref{P(e<H)}), with
$\theta_1=0$ and $\theta_0$ given in (\ref{eq:pen-theta0}) with $\Dth$ replaced by $\pi$ 
(see Fig.~\ref{fig:pen-integ-wb}a).

Integrating over $p$ and using (\ref{eq:energy}), it is straightforward to express,
without any approximation, the marginal in $\theta$ as
\begin{equation}
\label{eq:marginaltheta}
P_\theta(\theta,H)=\sqrt{2} \int_{-H\cos\theta}^{\infty} d \epsilon ~
\frac{1}{\sqrt{(\epsilon+H \cos \theta)}}~\bar{P}_\epsilon(\epsilon)~.
\end{equation}

The interpretation of $\bar P_\epsilon$ is straightforward. For any
given energy $\epsilon$, it is proportional to the ratio of the time taken to cover
the orbit in the $(\theta,p)$-plane confined by the waterbag domain to the time 
corresponding to the orbit unrestricted by the initial condition. 
$\bar P_\epsilon$ vanishes for $\epsilon < -H$, becomes $1/(4 \Dth \Dp)$
in the interval $-H < \epsilon < \epsilon^*$, with $\epsilon^*=\min (-H \cos \Dth, \Dp^2/2-H)$,
and monotonously decreases to zero in $\epsilon^* < \epsilon < \Dp^2/2 -H \cos \Dth$.

\section{Application to waterbag states of the HMF model}
\label{sec:HMF}

We now use these results in order to probe the steady states of the
HMF model, where the particles are now interacting. For the
distribution (\ref{eq:distr1}) to be a stationary state of the
corresponding HMF model one has to demand that the steady state
magnetization of the model with non interacting particles satisfies $H=m$
\cite{bouchet_variational}, where
\begin{equation}
\label{eq:self}
m=\int_{-\pi}^{+\pi} d\theta P_{\theta}(\theta,H) \cos \theta~.
\end{equation}
The strategy we propose is to solve the above self-consistency
equation in $H$ and then substitute the selected value of $H$ into
Eq.~(\ref{eq:distr1}) to get the steady state distributions, not only
of the uncoupled system but also of the HMF. This is the main result of this paper.
In applying these results to the HMF model, it is assumed that
the magnetization takes its steady state value instantaneously at
$t=0$. This is of course not correct since the magnetization evolves towards its steady
state value on a finite time scale, during which the momentum distribution also
changes. Thus the correspondence between the initial state and the steady
state distribution is only approximate.This approximations works well for some 
initial conditions and not so well for others, as discussed below. 

We proceed by solving the self-consistency equation (\ref{eq:self}). 
For all $\Dth < \pi$ and at small $H$ the leading term in the marginal
$P_{\theta}(\theta,H)$ behaves like $\sqrt{H}$. This can be easily proven numerically
but also analytically by splitting the integral in (\ref{eq:marginaltheta})
in the three relevant domains of integration $[-H \cos \theta,-H \cos \Dth]$,
$[-H \cos \Dth,\Dp^2/2-H]$, $[\Dp^2/2-H,\Dp^2/2-H \cos \Dth]$. The integration on both the first
and the second domain displays a $\sqrt{H}$ behavior and the first integral
can be explicitly computed, giving $2\sqrt{2H(\cos \theta - \cos \Dth)}$. 
The third integral gives an order $H^2$ contribution. When computing $m$ in formula (\ref{eq:self})
the first and second integral contribute terms that are opposite in sign for
large enough $\Dth$ and do not cancel each other. When $\Dth$ reaches $\pi$ the $\sqrt{H}$ 
terms cancel, giving rise in this limit to an order $H$ global contribution (we give below 
explicitly the result for $\Dth=\pi$). 
Because of the $\sqrt{H}$ behavior, the $H=0$ solution is always unstable. Therefore the 
magnetization is
expected to be non zero at any value of $\Dp$ \cite{antoniazzi_et_al_prl_2007,firpo}.
The theoretical magnetic curve obtained by (\ref{eq:marginaltheta}) and (\ref{eq:self}) is 
compared with the numerical
solution of the Vlasov equation for $\Dth=1$ in Fig.~\ref{fig:pen-integ-wb}b. The agreement 
is quite good, considering that the theoretical prediction has no free parameters.

\begin{figure}[ht]
 \centering
 \includegraphics[width=\linewidth]{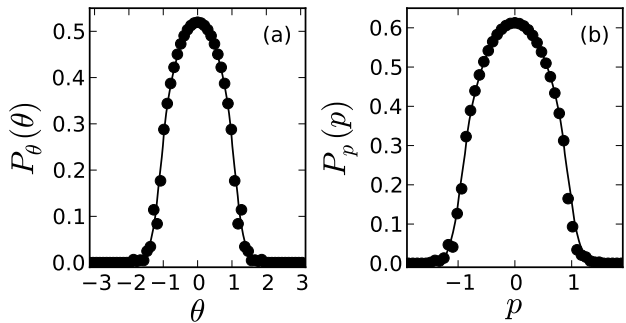}
 \caption{Marginals in $\theta$, $P_\theta(\theta)=\int dp ~ P(\theta,p)$,
   (panel a) and in $p$, $P_p(p)=\int d\theta ~ P(\theta,p)$, (panel b) of the
   steady state distribution (\ref{eq:distr1}) for $\Delta \theta=1$ and
   $\Delta p=1$. The full lines are the theoretical predictions and the points 
   are the results of numerical integrations of the Vlasov equation for
   the HMF model in Eqs.~(\ref{vlasov}) and (\ref{potential}).}
 \label{fig:marginals}
\end{figure}

In Fig.~\ref{fig:marginals} we compare the marginals in $\theta$ and in $p$ 
of the steady state distribution (\ref{eq:distr1}) with those obtained numerically by direct
integration of the Vlasov equation. Apart from the oscillations observed in the 
tails, which are due to the filamentation of the initial waterbag distribution, the
agreement is quite good.

\begin{figure}[ht]
  \centering
  \includegraphics[width=\linewidth]{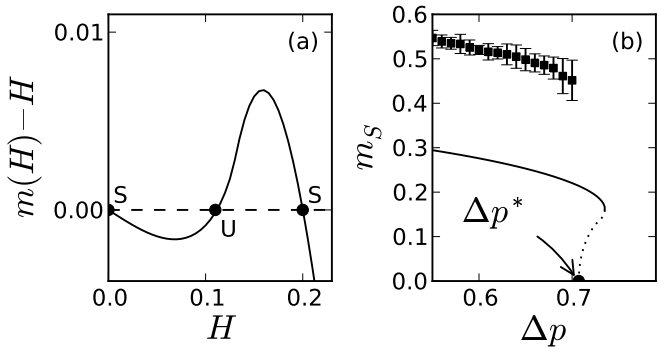}
  \caption{a) $m(H)-H$ vs. $H$ for $\Dp=0.72$ within the range of existence
  of three solutions $1/\sqrt{2}=\Dp^*<\Dp<\bar{\Dp}\simeq 0.735$. Two stable fixed point (S) with negative
  slopes and an unstable one (U) with a positive slope. b) The non zero steady state magnetization
  $m_S$ as a function of $\Dp$. The dotted segment corresponds to unstable solutions and the
  full line to stable ones. The dot on the real axis marks the stability limit of the homogeneous state 
  $\Dp^*$. The points are results of numerical integrations of the Vlasov equation for the HMF model in 
  Eqs.~(\ref{vlasov}) and (\ref{potential}).}
  \label{fig:consistency equation}
\end{figure}

Let us now consider initial homogeneous waterbag states, for which $\Dth=\pi$.
In analysing Eq.~(\ref{eq:self}) we find that $m=H=0$ is a solution
for any $\Dp$. At low $\Dp$ there exists one additional solution
with $m \neq 0$, while in a rather narrow intermediate range
$\Dp^*<\Dp<\bar{\Dp}$ two solutions with nonzero magnetization are present. In
Fig.~\ref{fig:consistency equation}a we plot $m(H)-H$ as a function of
$H$ for $\Dp$ in the intermediate region where one gets three
solutions of the self-consistency equation, two stable (S) and one unstable (U).
In Fig.~\ref{fig:consistency equation}b we plot the non zero self-consistent
steady value of the magnetization, $m_S$, versus $\Dp$. In this case the
theoretical prediction does not agree with the numerical simulations of the
Vlasov equation. We argue that this is due to the fact that, during time
evolution, there is a significant mixing of energy levels, due to strong
variations of the effective field near a discontinuous first order transition
\cite{pakter}.

\section{Stability of the homogeneous waterbag state}
\label{sec:stability}

We now analyze the stability of the stationary solutions.
For the zero magnetization state the stability limit is determined
by the slope of $m(H)$ at $H=0$, which is obtained by taking
the small $H$ limit in the integral (\ref{eq:marginaltheta}).
We therefore expand $P_\theta(\theta,H)$
in powers of $H$ keeping only the linear term in $H$ which is proportional
to $\cos \theta$. This term is found to be $H\cos \theta /(2 \pi \Dp^2)$. 
It is obtained by carrying out
the integration in the energy domain $-H \cos \theta <\epsilon <
\Delta p^2/2-H$, where $\bar P_\epsilon(\epsilon)$  is constant.
The stability threshold for the $m=0$ solution is thus
$\Dp^*=1/\sqrt{2}$, as displayed in Fig.\ref{fig:consistency equation}b. 
For $\Dp > \Dp^*$ the homogeneous state is stable and it
becomes unstable below $\Dp^*$. Using this value in the HMF model we
get the energy $(\Dp^*)^2/6+1/2=7/12$, which coincides with the
threshold energy calculated using other methods \cite{yamaguchi_et_al_physica_a_2004,HMF}.
This analysis can be generalized to a generic initial momentum
distribution.

\begin{figure}[ht]
  \centering
  \includegraphics[width=\linewidth]{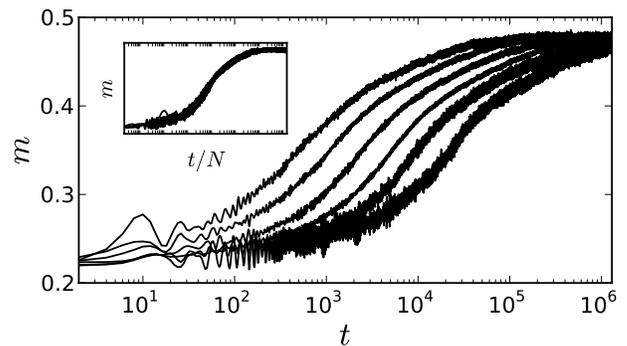}
  \caption{Magnetization vs. time of the HMF model starting with the stable inhomogeneous
  steady state at $\Dp=0.7$ for $N=2^{10}(100),2^{11}(100),2^{12}(50),2^{13}(50),2^{14}(20),2^{15}(20)$ from left 
  to right (in parenthesis the number of realisations). In the inset the same data are plotted
  as a function of $t/N$.}
  \label{fig:scaling}
\end{figure}

The stable inhomomogeneous solutions found for different values of $\Dp$
are exact stable stationary solutions of the Vlasov equation, but they evolve
in time for finite $N$. 
Numerical studies of the time evolution of the magnetization are summarized
in Fig.~\ref{fig:scaling}, where $m(t)$ is displayed for several values of $N$.
We obtain a data collapse when time is rescaled by $1/N$, implying that
the relaxation time increases as $N^\gamma$ with $\gamma=1$ \cite{chavanis,joyce}.
This is at variance with the relaxation of homogeneous states where
$\gamma \simeq 1.7$ \cite{yamaguchi_et_al_physica_a_2004,anteneodo,gupta}.

\section{Conclusions}
\label{sec:conclusions}

In summary, our approach allows one to derive exact inhomogeneous steady states.
It gives approximate correspondence between initial and 
final states, which, for the HMF model, yields good results
for initial inhomogeneous waterbags. Moreover, one obtains the exact
stability limit of the zero magnetization state, since
the waterbag initial condition is in itself a steady state with zero
magnetization and the dynamics of each particle is therefore exactly decribed 
by the independent particles model (\ref{eq:energy}).

\section{Acknowledgements}
We thank O. Cohen, O. Hirschberg, S. Gupta and V. Rittenberg for helpful
discussions. The support of the Israel Science Foundation (ISF), the Minerva
Foundation, with funding from the Federal German Ministry for Education and Reseach, INFN 
and the Belgian federal government (IAP Project ``NOSY'' P6/02) is gratefully
acknowledged. This work was carried out while S.R. was Weston Visiting Professor at
the Weizmann Institute of Science and is also part of the ANR-10-CEXC-010-01, {\it Chaire 
d'Excellence} project.


\begin{thebibliography}{99}


\bibitem{reviews}
T. Dauxois, S. Ruffo and L. Cugliandolo (Eds.), {\it Long-Range Interacting Systems}
(Oxford University Press, (2009);
F. Bouchet, S. Gupta and D. Mukamel, Physica A {\bf 389}, 4389 (2010);
A. Campa, T. Dauxois and S. Ruffo,
Phys. Rep. {\bf 480}, 57 (2009).

\bibitem{yamaguchi_et_al_physica_a_2004}
Y.~Y. Yamaguchi, J. Barr{\'e}, F. Bouchet, T. Dauxois and S. Ruffo,
Physica A {\bf 337}, 36 (2004).

\bibitem{HMF}
M. Antoni and S. Ruffo,
Phys. Rev. E {\bf 52}, 2361 (1995);P. H. Chavanis, J. Vatteville and F. Bouchet,
European Physical Journal {\bf B46}, 61 (2005).


\bibitem{nicholson}
D. R. Nicholson, {\em Introduction to Plasma Theory},
(John Wiley, New York, 1983).

\bibitem{inhomogeneous}
K. Jain, F. Bouchet and D. Mukamel,
J. Stat. Mech. {\bf 11}, 8 (2010);
J. Barr{\'e}, A. Olivetti and Y. Y. Yamaguchi,
J. Stat. Mech. P08002 (2010);
A. Campa and P.-H. Chavanis, J. Stat. Mech. P06001 (2010);
R. Bachelard, F. Staniscia, T. Dauxois, G. De Ninno and S. Ruffo,
J. Stat. Mech., P03022 (2011).

\bibitem{BGK}
I. Bernstein, J. M. Greene, M. D. Kruskal 
Phys. Rev., {\bf 108}, 546 (1957).

\bibitem{Yamaguchi_2011}
Y. Y. Yamaguchi, Phys. Rev. E, {\bf 84}, 016211 (2011).

\bibitem{pomeau_2007}
Y. Pomeau, {\it Statistical mechanics of gravitational plasmas},
Lecture Notes, 2nd Warsaw School of Statistical Physics  (2007).

\bibitem{leoncini_et_al_epl_2009}
X. Leoncini, T.~Van~Den Berg and D. Fanelli,
Europhys. Lett. {\bf 86}, 20002 (2009).

\bibitem{de_buyl_mukamel_ruffo_inprep}
P. de~Buyl, D. Mukamel and S. Ruffo,
Phil. Trans. R. Soc. A, {\bf 369}, 439 (2011).

\bibitem{lynden-bell_1967}
D.~Lynden-Bell,
Mon. Not. R. Astron. Soc. {\bf 136}, 101 (1967).


\bibitem{barre_et_al_pre_2004}
J. Barr{\'e}, T. Dauxois, G. De Ninno, D. Fanelli and S. Ruffo,
Phys. Rev. E {\bf 69}, 045501 (2004).


\bibitem{antoniazzi_et_al_prl_2007}
A. Antoniazzi, D. Fanelli, S. Ruffo, and Y.~Y. Yamaguchi,
Phys. Rev. Lett. {\bf 99}, 040601 (2007).

\bibitem{levin}
Y. Levin, R. Pakter and T.~N. Teles,
Phys. Rev. Lett. {\bf 100}, 040604 (2008); T.N. Teles, Y.Levin, R. Pakter and F.B. Rizzato, 
J. Stat. Mech. P05007 (2010). 

\bibitem{bouchet_variational}
F. Bouchet, 
Physica D, {\bf 237}, 1976 (2008).

\bibitem{firpo}
M.-C. Firpo,
Europhys. Lett. {\bf 88}, 30010 (2009).

\bibitem{pakter}
R. Pakter and Y. Levin,
Phys. Rev. Lett. {\bf 106}, 200603 (2011).

\bibitem{chavanis}
P.-H. Chavanis
J. Stat. Mech., P05019 (2010).

\bibitem{joyce}
M Joyce, T Worrakitpoonpon
J. Stat. Mech, P10012 (2010).


\bibitem{anteneodo}
L. G. Moyano and C. Anteneodo,
Phys. Rev. E {\bf 74}, 021118 (2006).

\bibitem{gupta}
S. Gupta and D. Mukamel,
Phys. Rev. Lett. {\bf 105}, 040602 (2010).





\end{thebibliography}
\end{document}